\documentclass{article}
\usepackage{graphicx}
\usepackage[english]{babel}

\usepackage{amsmath}
\usepackage{hyperref}
\usepackage{color}
\usepackage{cite}

\title{
\includegraphics[width=0.35\textwidth]{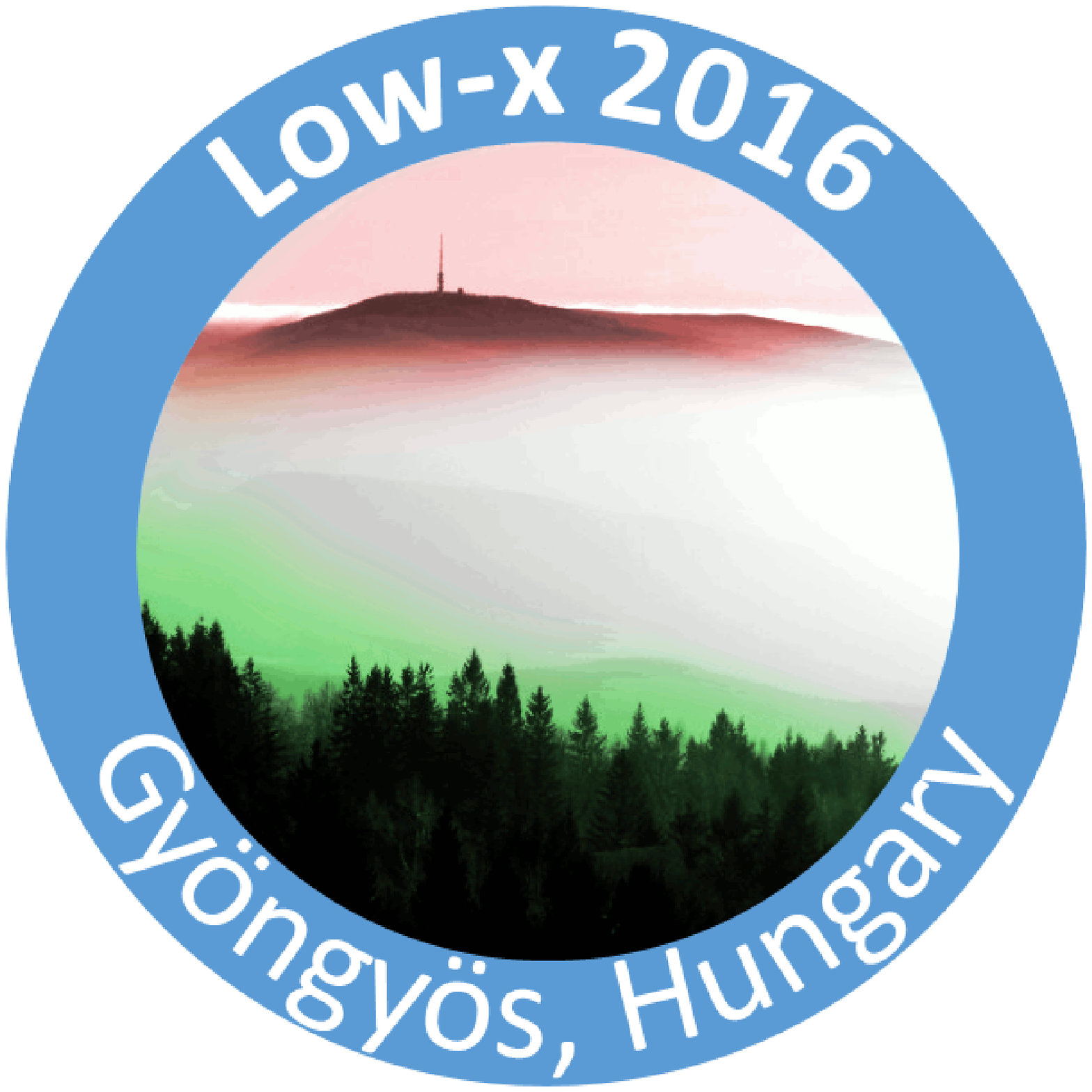}\\[1cm]
Diffractive processes at the LHC \\ within kt -factorization approach
}
\author{{Marta {\L}uszczak$^1$, Antoni Szczurek$^{1,2}$}
\\[1ex]
$^1$University of Rzesz\'ow, PL-35-959 Rzesz\'ow, Poland\\
$^2$Institute of Nuclear Physics PAN, PL-31-342 Krak{\'o}w, Poland\\
}

\begin{document}

\fontfamily{lmss}\selectfont
\maketitle

\begin{abstract}
We discuss the single diffractive production of $c \bar c$ pairs and charmed mesons at the LHC. In addition to standard collinear approach, for a first time we propose a $k_t$-factorization approach to the diffractive processes. The transverse momentum dependent (the unintegrated diffractive parton distributions) in proton are obtained with the help of the Kimber-Martin-Ryskin
prescription where collinear diffractive PDFs are used as input.
In this calculation the transverse momentum of the pomeron is neglected  with respect to transverse momentum of partons entering the hard process. The results of the new approach are compared with those of the standard collinear one. Significantly larger cross sections are obtained in the $k_t$-factorization approach where some part of higher-order effects is effectively included. Some correlation observables, like azimuthal angle correlation between $c$ and $\bar c$, and $c \bar c$ pair transverse momentum distribution were obtained for the first time.
\end{abstract}

\section{Introduction}
Diffractive hadronic processes  were studied theoretically in the
so-called resolved pomeron model \cite{IS}. This model, previously
used to describe deep-inelastic diffractive processes 
must be corrected for absorption effects related to
hadron-hadron interactions. 
In theoretical models this effect is taken into account approximately 
by multiplying the diffractive cross section calculated using HERA diffractive PDFs 
by a kinematics independent factor called the gap survival probability -- $S_{G}$. 
Two theoretical groups specialize in calculating such probabilities
\cite{KMR2000, Gotsman:2005rt}.

In this study we consider diffractive production of charm for which
rather large cross section at the LHC are expected, 
even within the leading-order (LO) collinear approach
\cite{Luszczak:2014cxa}. 
On the other hand, it was shown that for the inclusive non-diffractive 
charm production the LO collinear approach is a rather poor
approximation and higher-order corrections are crucial. 
Contrary, the $k_t$-factorization approach, which effectively includes 
higher-order effects, gives a good description of the LHC data  
for inclusive charm production at $\sqrt{s}$ = 7 TeV 
(see \textit{e.g}. Ref.~\cite{Maciula:2013wg}). 
This strongly suggests that application of the $k_t$-factorization approach 
to diffractive charm production is useful. 
Besides, the dipole approach is also often used to calculate cross section  for diffractive processes. However, as we discussed in Ref.~\cite{Luszczak:2013cba}, it gives only a small fraction of the diffractive cross section for the charm production.
This presentation is based on our recent study presented in \cite{Luszczak:2016csq}.
Here we present only results at the quark/antiquark level.

\section{A sketch of the theoretical formalism}

\begin{figure}[!htbp]
\begin{minipage}{0.4\textwidth}
 \centerline{\includegraphics[width=1.0\textwidth]{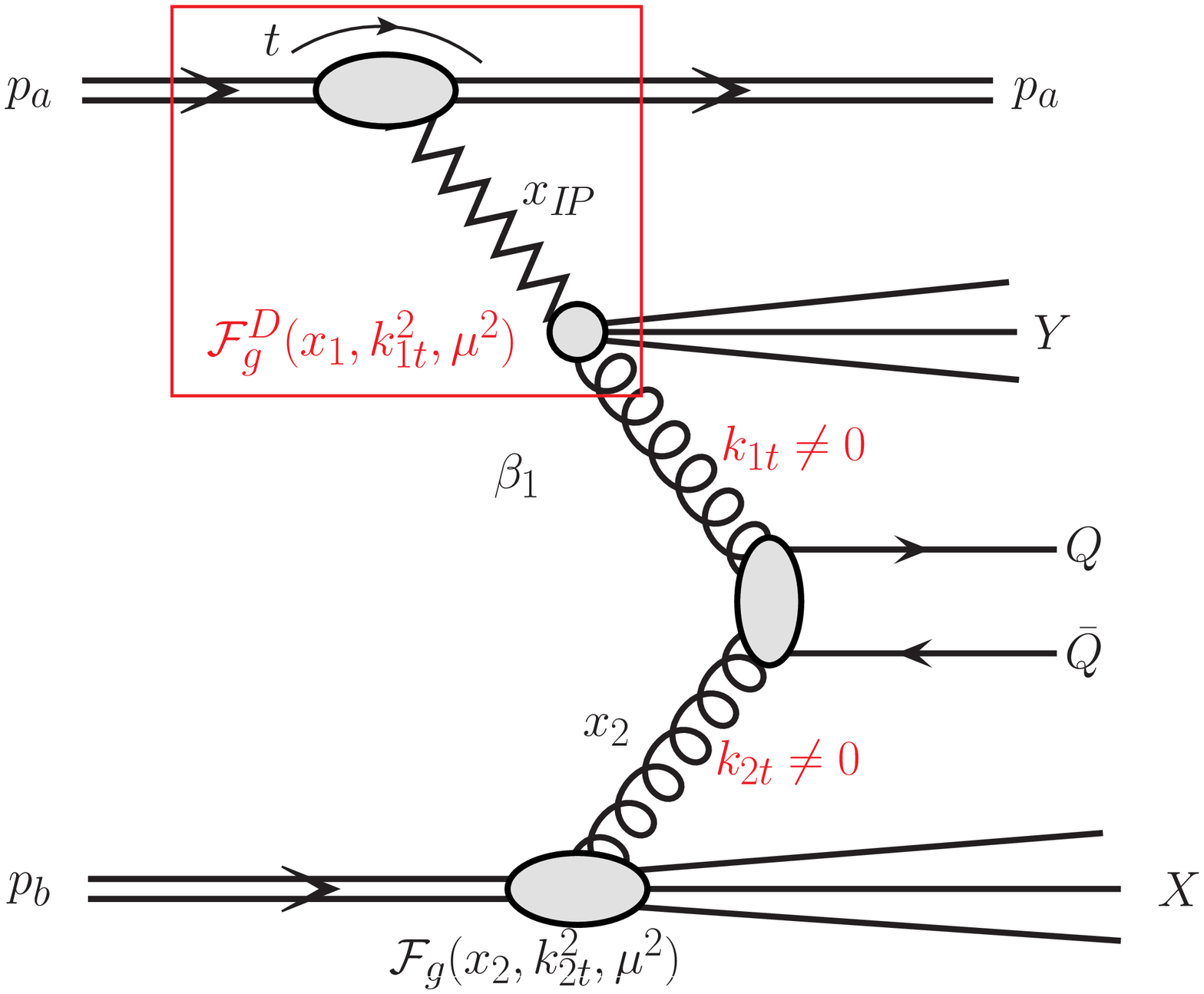}}
\end{minipage}
\begin{minipage}{0.4\textwidth}
 \centerline{\includegraphics[width=1.0\textwidth]{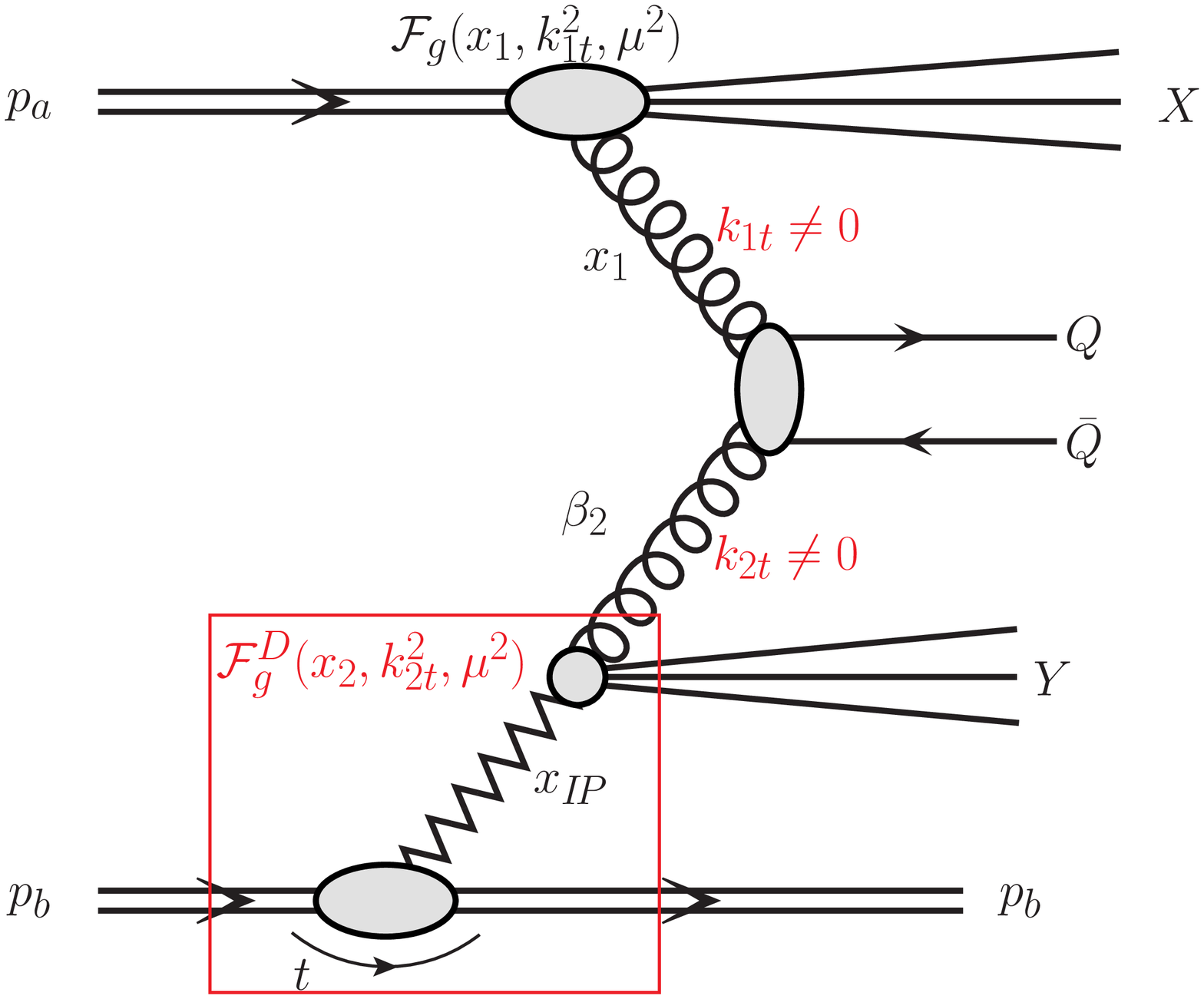}}
\end{minipage}
\caption{
\small A diagrammatic representation for single-diffractive production
of heavy quark pairs within the $k_{t}$-factorization resolved pomeron approach.
}
 \label{fig:mechanism}
\end{figure}

A sketch of the theoretical formalism is shown in
Fig.~\ref{fig:mechanism}. Here, extension of the standard resolved
pomeron model based on the LO collinear approach by adopting a framework
of the $k_{t}$-factorization is proposed as an effective way to include higher-order corrections.
According to this model the cross section for a single-diffractive production of charm quark-antiquark pair, for both considered diagrams (left and right panel of Fig.~\ref{fig:mechanism}), can be written as:
\begin{eqnarray}
d \sigma^{SD(a)}({p_{a} p_{b} \to p_{a} c \bar c \; X Y}) &=&
\int d x_1 \frac{d^2 k_{1t}}{\pi} d x_2 \frac{d^2 k_{2t}}{\pi} \; d {\hat \sigma}({g^{*}g^{*} \to c \bar c }) \nonumber \\
&& \times \; {\cal F}_{g}^{D}(x_1,k_{1t}^2,\mu^2) \cdot {\cal F}_{g}(x_2,k_{2t}^2,\mu^2) ,
\label{SDA_formula}
\end{eqnarray}
\begin{eqnarray}
d \sigma^{SD(b)}({p_{a} p_{b} \to c \bar c p_{b} \; X Y}) &=&
\int d x_1 \frac{d^2 k_{1t}}{\pi} d x_2 \frac{d^2 k_{2t}}{\pi} \; d {\hat \sigma}({g^{*}g^{*} \to c \bar c }) \nonumber \\
&& \times \; {\cal F}_{g}(x_1,k_{1t}^2,\mu^2) \cdot {\cal F}_{g}^{D}(x_2,k_{2t}^2,\mu^2),
\label{SDB_formula}
\end{eqnarray}
where ${\cal F}_{g}(x,k_{t}^2,\mu^2)$ are the "conventional" unintegrated ($k_{t}$-dependent) gluon distributions (UGDFs) in the proton and ${\cal F}_{g}^{D}(x,k_{t}^2,\mu^2)$ are their diffractive counterparts. The latter can be interpreted as a probability of finding a gluon with longitudinal momentum fraction $x$ and transverse momentums $k_{t}$ at the factorization scale $\mu^{2}$ assuming that the proton which lost a momentum fraction $x_{I\!P}$ remains intact.

Details of our new calculations can be found in Ref.~\cite{Luszczak:2016csq}.

\section{Selected results}
\begin{figure}[!htbp]
\begin{minipage}{0.47\textwidth}
 \centerline{\includegraphics[width=1.0\textwidth]{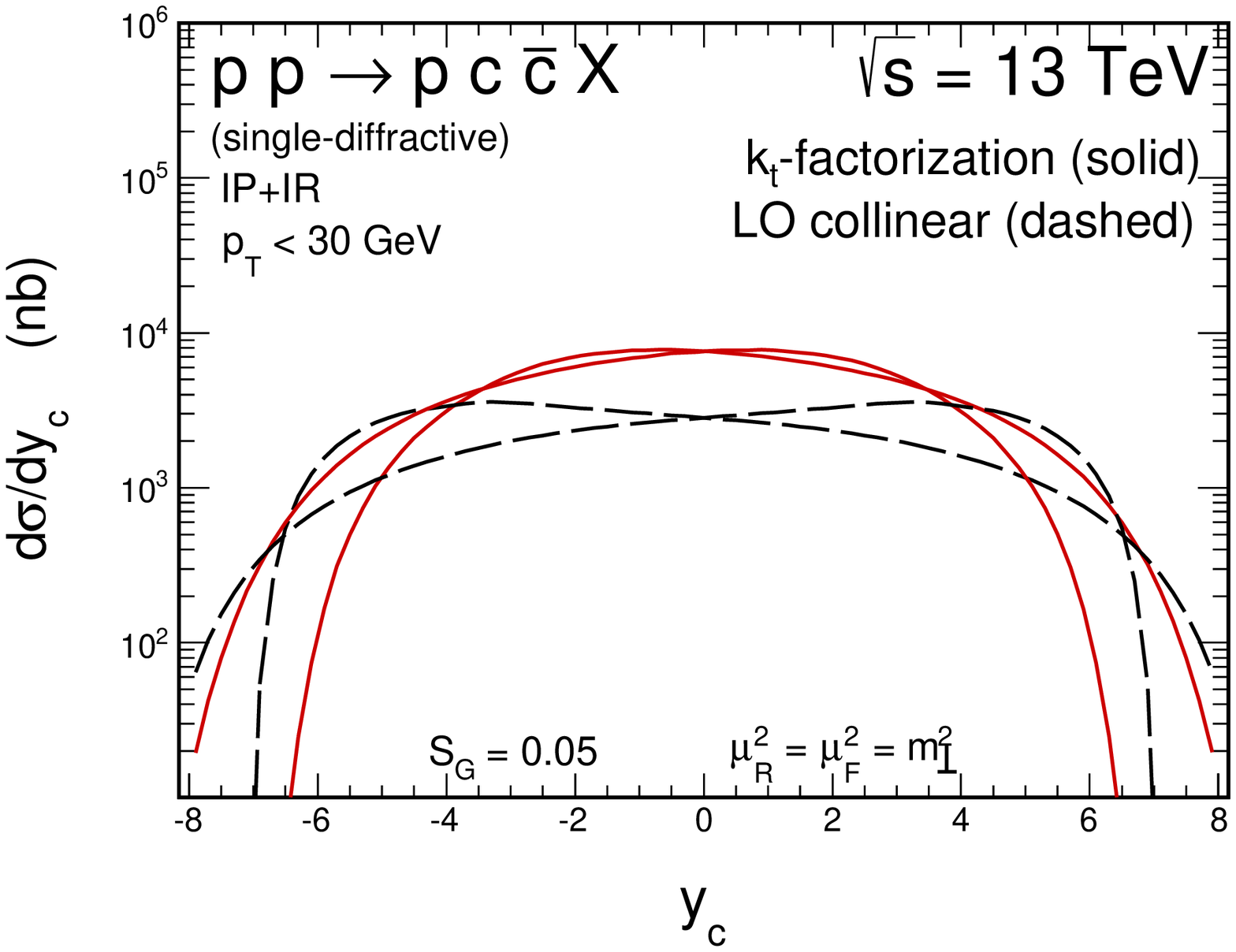}}
\end{minipage}
\hspace{0.5cm}
\begin{minipage}{0.47\textwidth}
 \centerline{\includegraphics[width=1.0\textwidth]{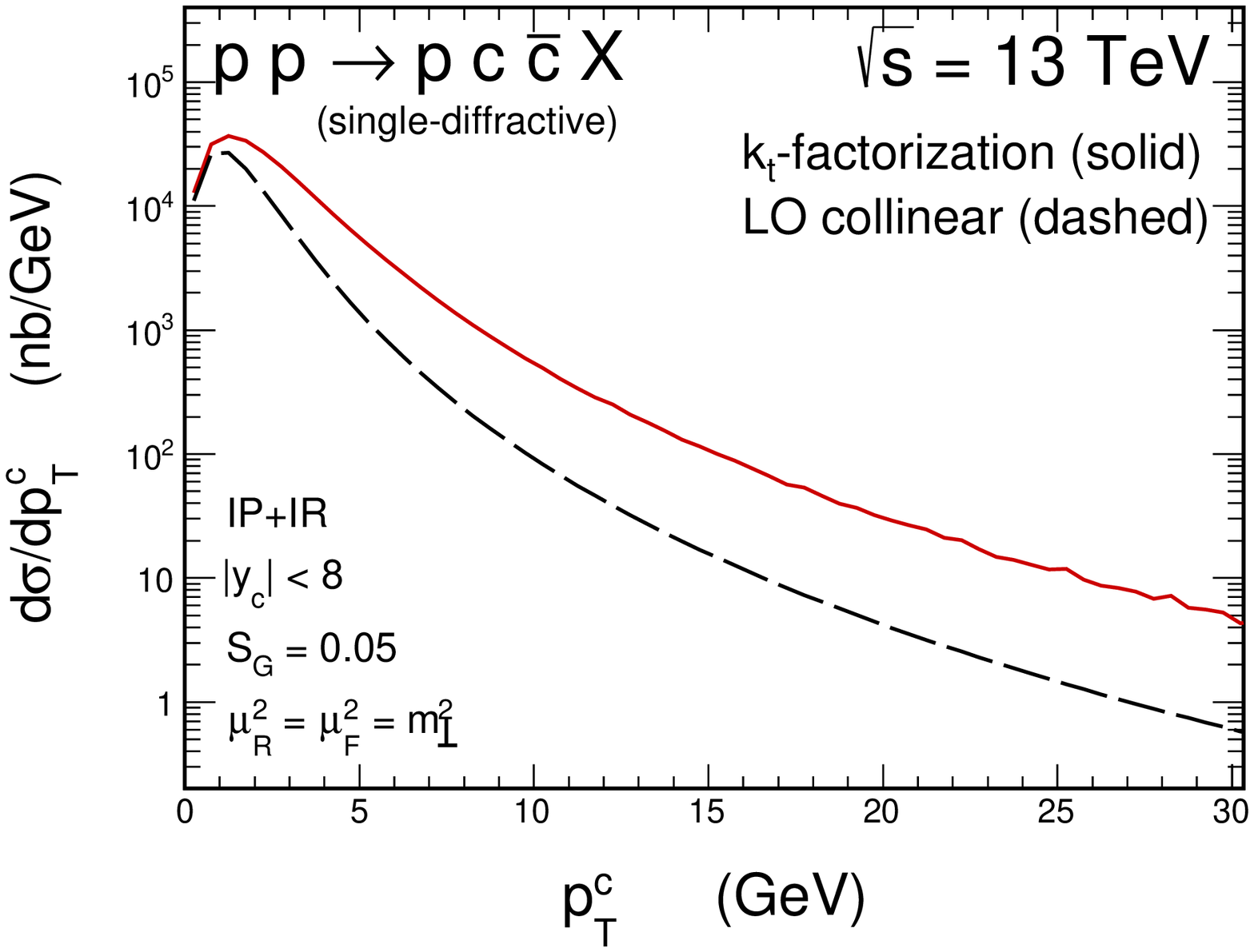}}
\end{minipage}
   \caption{
\small Rapidity (left panel) and transverse momentum (right panel) distributions of $c$ quarks (antiquarks) for a single-diffractive production at $\sqrt{s} = 13$ TeV. Components of the $g(I\!P)\operatorname{-}g(p)$, $g(p)\operatorname{-}g(I\!P)$, $g(I\!R)\operatorname{-}g(p)$, $g(p)\operatorname{-}g(I\!R)$ mechanisms are shown. 
}
 \label{fig:ypt_kTcoll}
\end{figure}

First, we show some selected examples of the results of the
$k_T$-factorization calculation in Fig.~\ref{fig:ypt_kTcoll}.
In Fig.~\ref{fig:ypt_kTcoll} we show rapidity (left panel) and
transverse momentum (right panel) distribution of $c$ quarks
(antiquarks) for single diffractive production at $\sqrt{s} = 13$ TeV.
Distributions calculated within the LO collinear factorization (black
long-dashed lines) and for the $k_{t}$-factorization approach (red solid
lines) are shown separately. We see significant differences between results of the both approaches, that are consistent with the conclusions from similar
studies of standard non-diffractive charm production (see
\textit{e.g}. Ref.~\cite{Maciula:2013wg}). 
Here we confirm that the higher-order corrections are very important
also for the diffractive production of charm quarks.

\begin{figure}[!htbp]
\begin{minipage}{0.47\textwidth}
 \centerline{\includegraphics[width=1.0\textwidth]{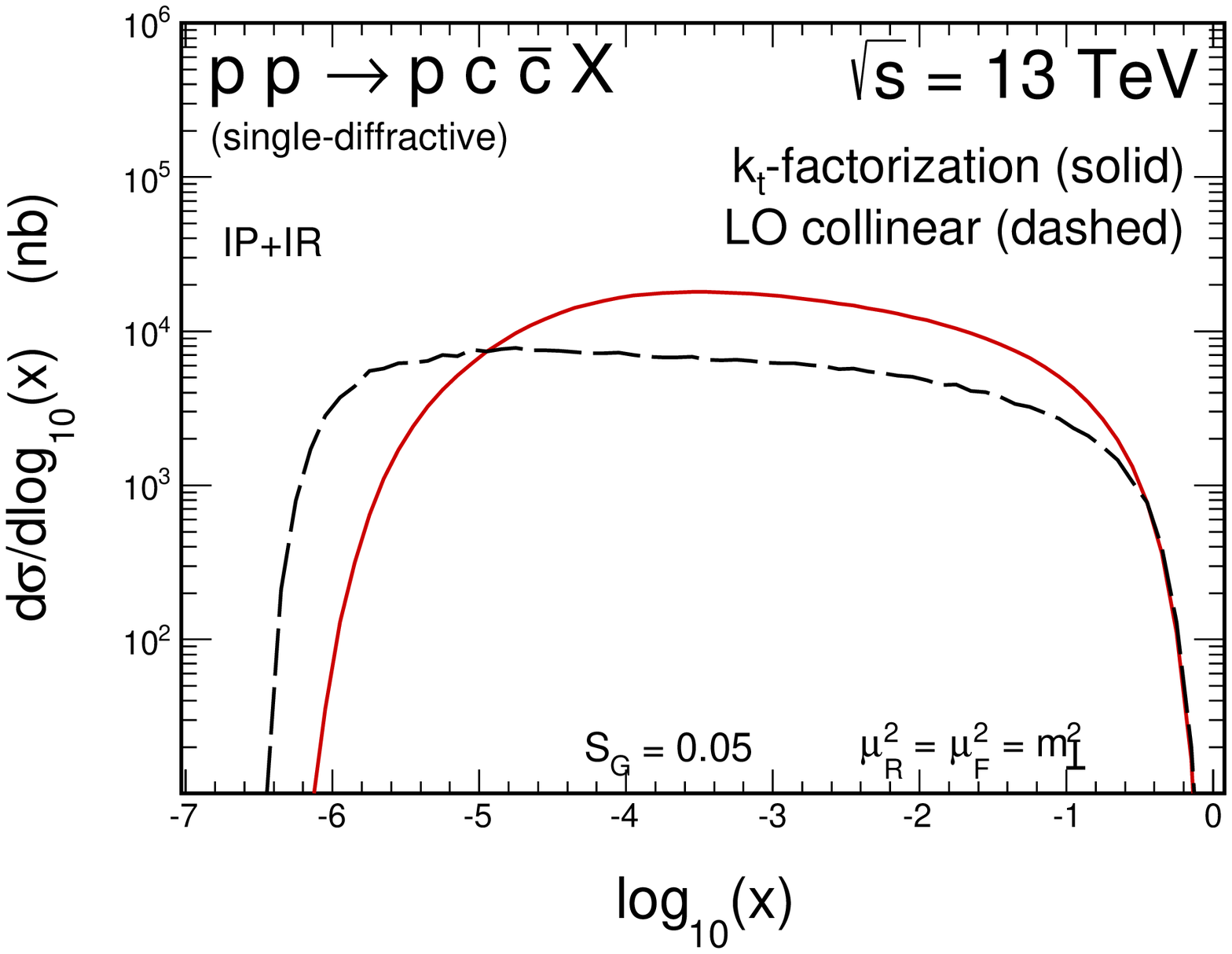}}
\end{minipage}
\hspace{0.5cm}
\begin{minipage}{0.47\textwidth}
 \centerline{\includegraphics[width=1.0\textwidth]{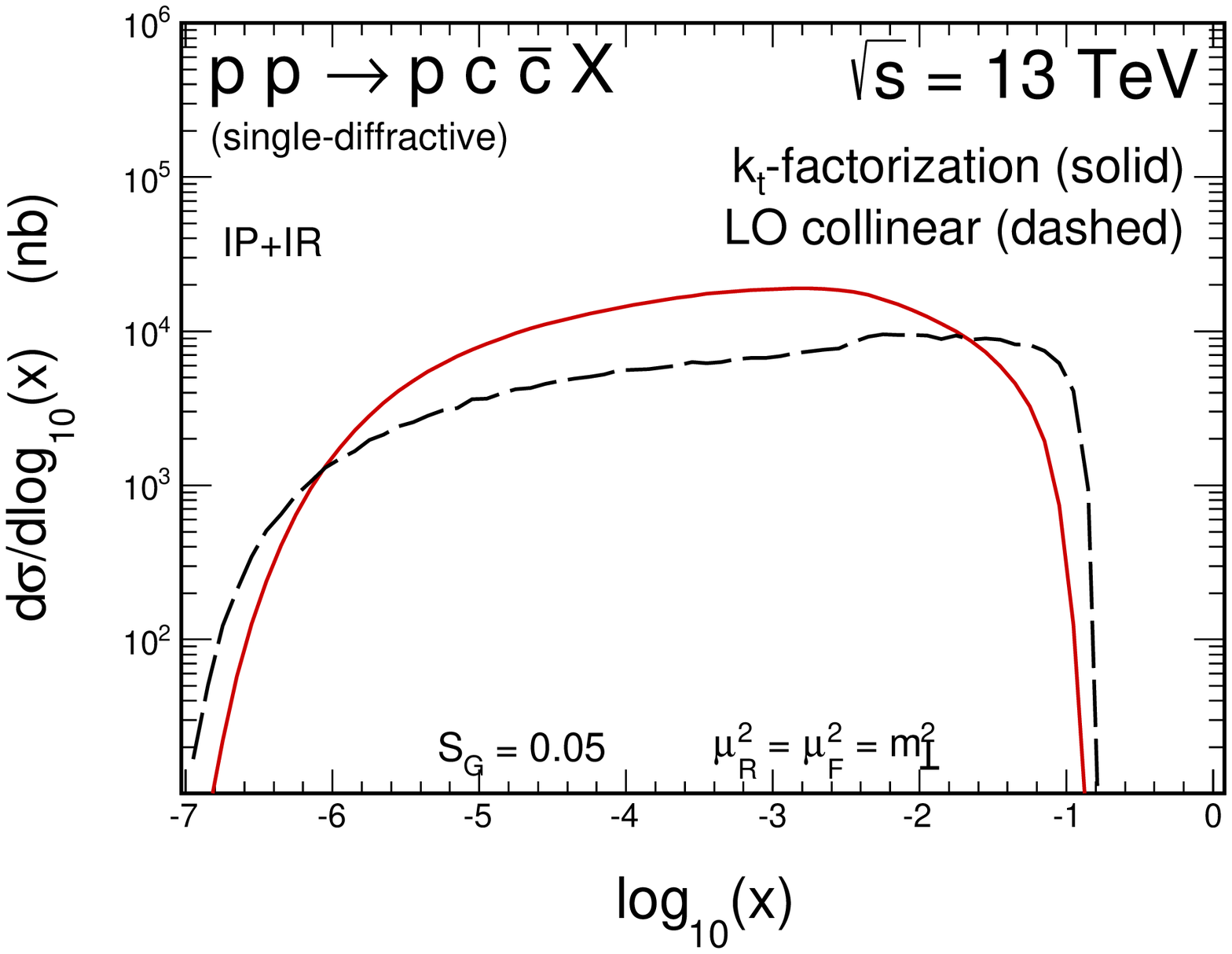}}
\end{minipage}
   \caption{
\small The differential cross section as a function of $\log_{10}(x)$ with $x$ being the non-diffractive gluon longitudinal momentum fraction (left panel) and the diffractive gluon longitudinal momentum fraction with respect to the proton (right panel) for single-diffractive production at $\sqrt{s} = 13$ TeV. Results for the LO collinear (black long-dashed) and the $k_{t}$-factorization (red solid) approaches are compared.
}
 \label{fig:logx_kTcoll}
\end{figure}

Figure ~\ref{fig:logx_kTcoll} shows the differential cross section as a function of $\log_{10}(x)$ where $x$ is defined as the longitudinal momentum fraction of proton carried by the gluon from non-diffractive side (left panel) or as the longitudinal momentum fraction of proton carried by the diffractive gluon emitted from pomeron/reggeon on diffractive side (right panel). In the case of non-diffractive gluon (left panel) we see that for extremely small values of $x$ the LO collinear predictions strongly exceed the ones of the $k_{t}$-factorization. This effect also affects the rapidity spectra in the very forward/backward regions (see Fig.~\ref{fig:ypt_kTcoll}) and is partially related to a very poor theoretical control of the collinear PDFs in the range of $x$ below $10^{-5}$.  

\begin{figure}[!htbp]
\begin{minipage}{0.47\textwidth}
 \centerline{\includegraphics[width=1.0\textwidth]{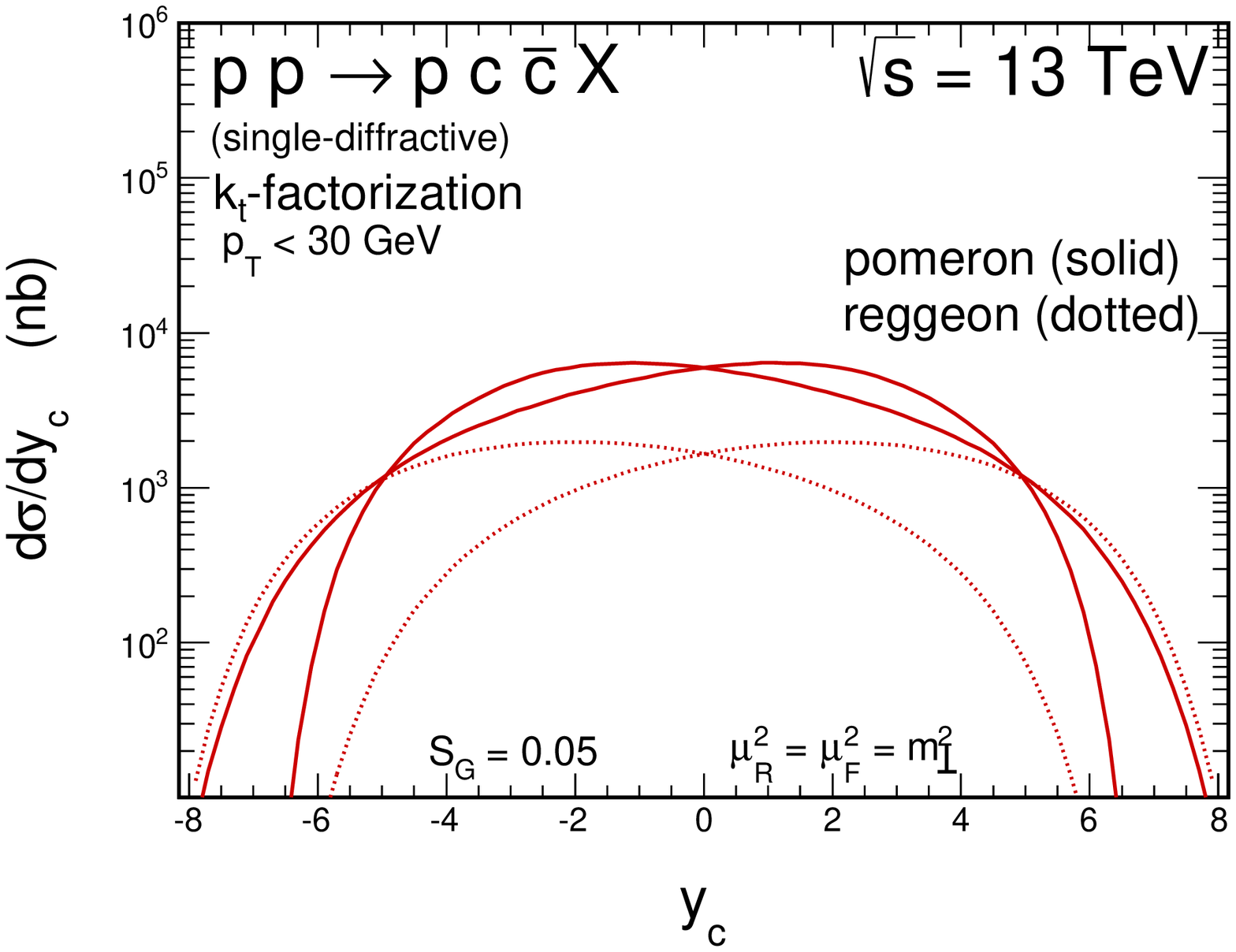}}
\end{minipage}
\hspace{0.5cm}
\begin{minipage}{0.47\textwidth}
 \centerline{\includegraphics[width=1.0\textwidth]{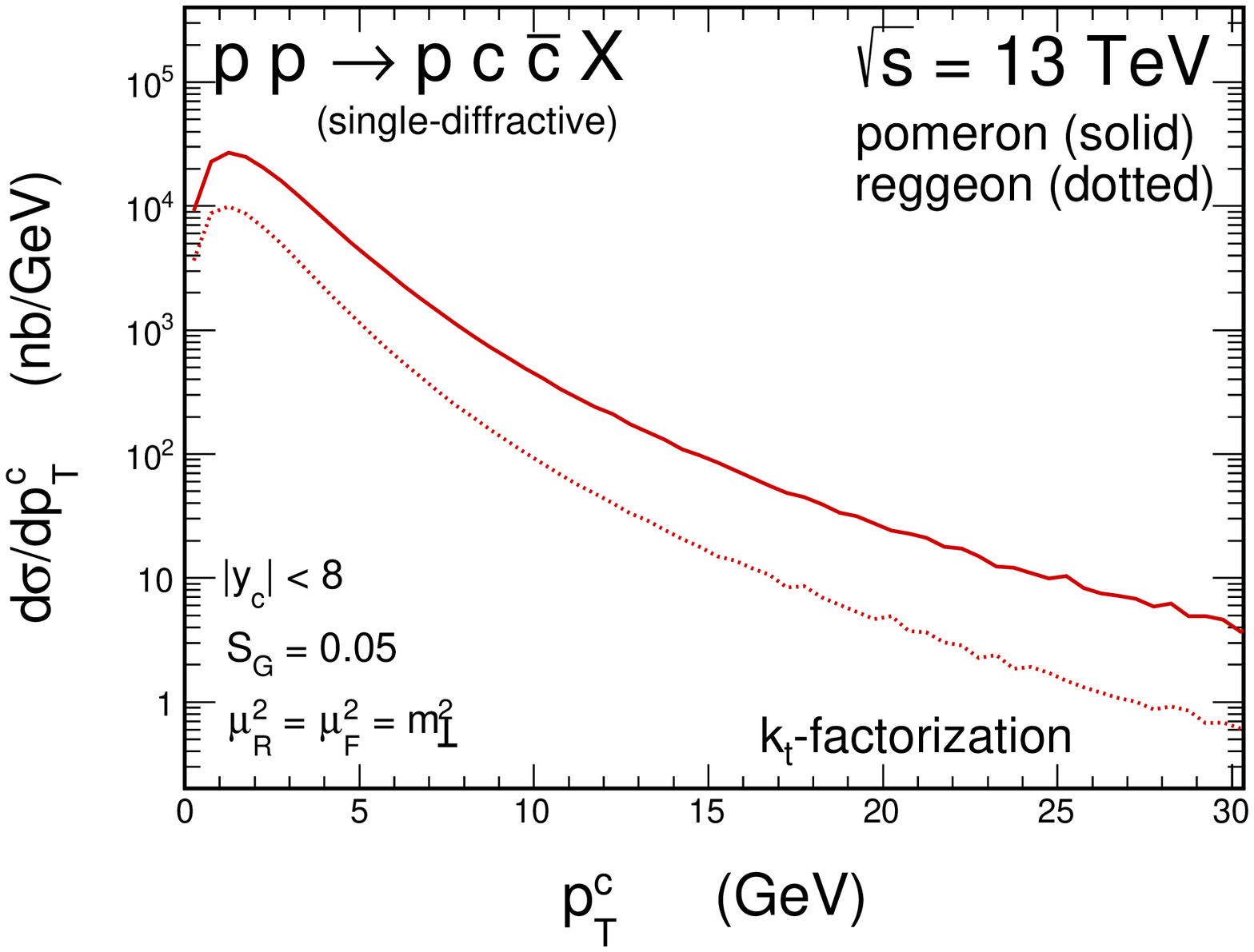}}
\end{minipage}
   \caption{
\small Rapidity (left panel) and transverse momentum (right panel) distributions of $c$ quarks (antiquarks) for single-diffractive production at $\sqrt{s} = 13$ TeV calculated with the $k_{t}$-factorization approach. Contributions of the $g(I\!P)\operatorname{-}g(p)$, $g(p)\operatorname{-}g(I\!P)$, $g(I\!R)\operatorname{-}g(p)$, $g(p)\operatorname{-}g(I\!R)$ mechanisms are shown separately.
}
 \label{fig:ypt_kT_PR}
\end{figure}

In Fig.~\ref{fig:ypt_kT_PR} we show again the rapidity (left panel) and transverse momentum (right panel) distributions of $c$ quarks (antiquarks) calculated in the $k_{t}$-factorization approach. Here contributions from the pomeron and the reggeon exchanges are shown separately. The estimated sub-leading reggeon contribution is of similar size as the one of the leading pomeron. In the single-diffractive case the maxima of rapidity distributions for $g(I\!P)\operatorname{-}g(p)$ and $g(p)\operatorname{-}g(I\!P)$ (or $g(I\!R)\operatorname{-}g(p)$ and $g(p)\operatorname{-}g(I\!R)$) mechanisms are shifted to forward and backward rapidities with respect to the non-diffractive case. This is related to the upper limit on diffractive gluon longitudinal momentum fraction ($x \leq x_{I\!P}$).  

\begin{figure}[!htbp]
\begin{minipage}{0.47\textwidth}
 \centerline{\includegraphics[width=1.0\textwidth]{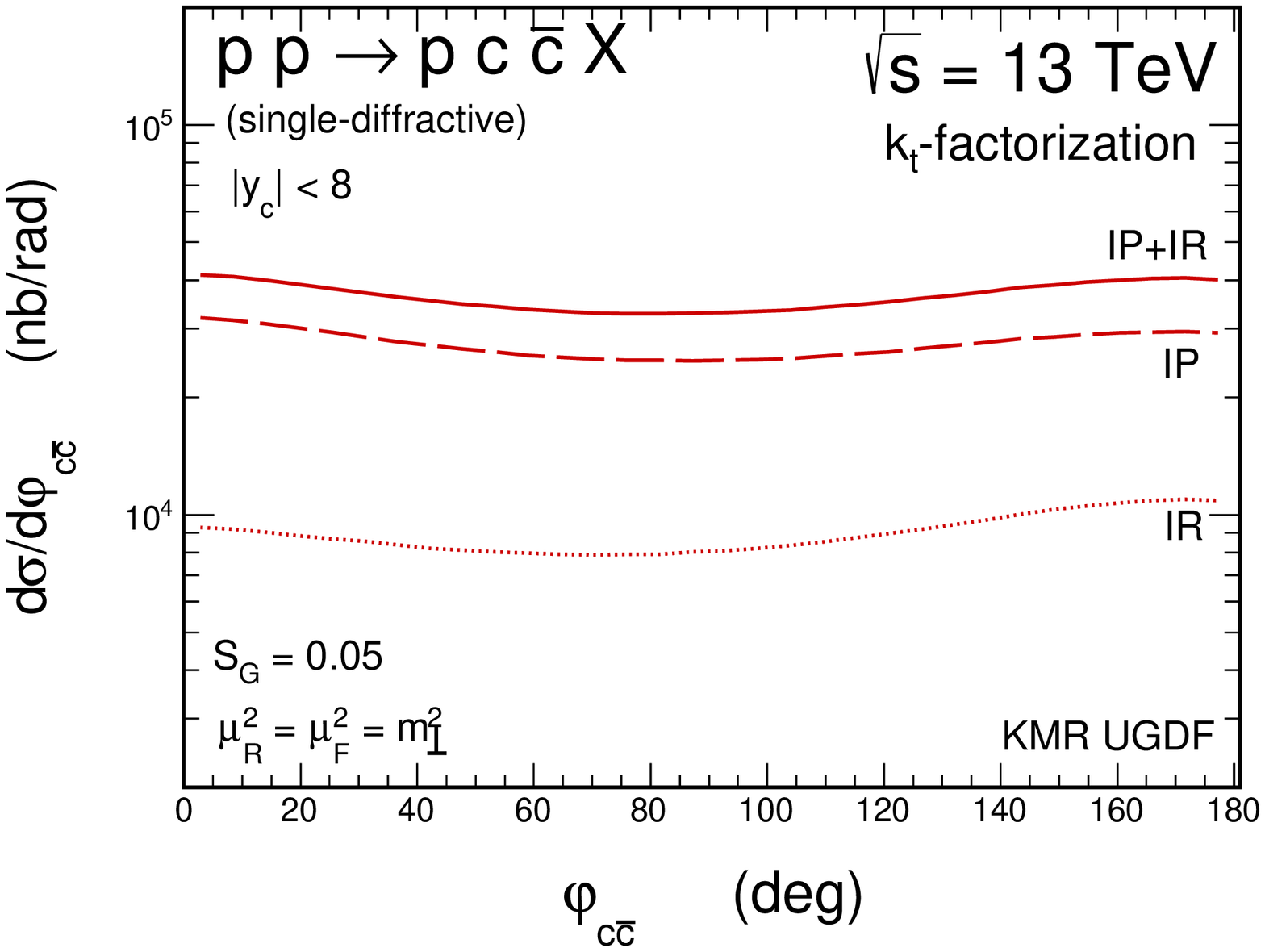}}
\end{minipage}
\hspace{0.5cm}
\begin{minipage}{0.47\textwidth}
 \centerline{\includegraphics[width=1.0\textwidth]{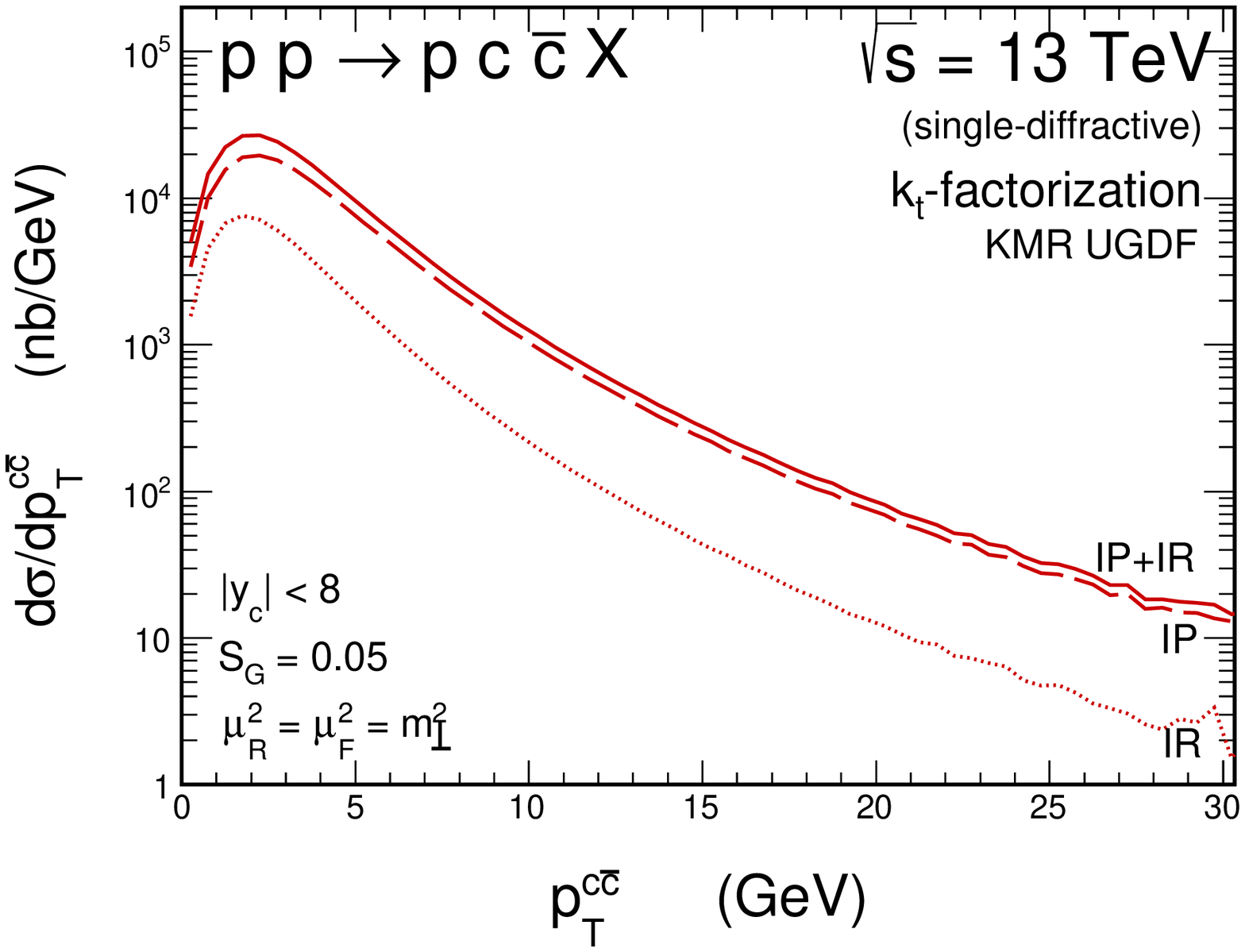}}
\end{minipage}
   \caption{
\small The distribution in $\phi_{c \bar c}$ (left panel) and
distribution in $p_{T}^{c\bar c}$ (right panel) in the
$k_{t}$-factorization approach at $\sqrt{s}$ = 13 TeV.
}
 \label{fig:phid_ptsum_kT_PR}
\end{figure}

The correlation observables cannot be calculated within the LO collinear
factorization but can be directly obtained in the $k_{t}$-factorization
approach.
The distribution of azimuthal angle $\varphi_{c \bar c}$ between $c$
quarks and $\bar c$ antiquarks is shown in the left panel of
Fig.~\ref{fig:phid_ptsum_kT_PR}. The $c \bar c$ pair transverse momentum
distribution 
$p^{c \bar c}_{T} = |\vec{p^{c}_{t}} + \vec{p^{\overline{c}}_{t}}|$ 
is shown in the right panel. Results of the full phase-space
calculations illustrate that the quarks and antiquarks
in the $c \bar c$ pair are almost uncorrelated in the azimuthal angle
between them and are often produced in the configuration with quite
large pair transverse momenta. 

\begin{figure}[!htbp]
\begin{minipage}{0.4\textwidth}
 \centerline{\includegraphics[width=1.0\textwidth]{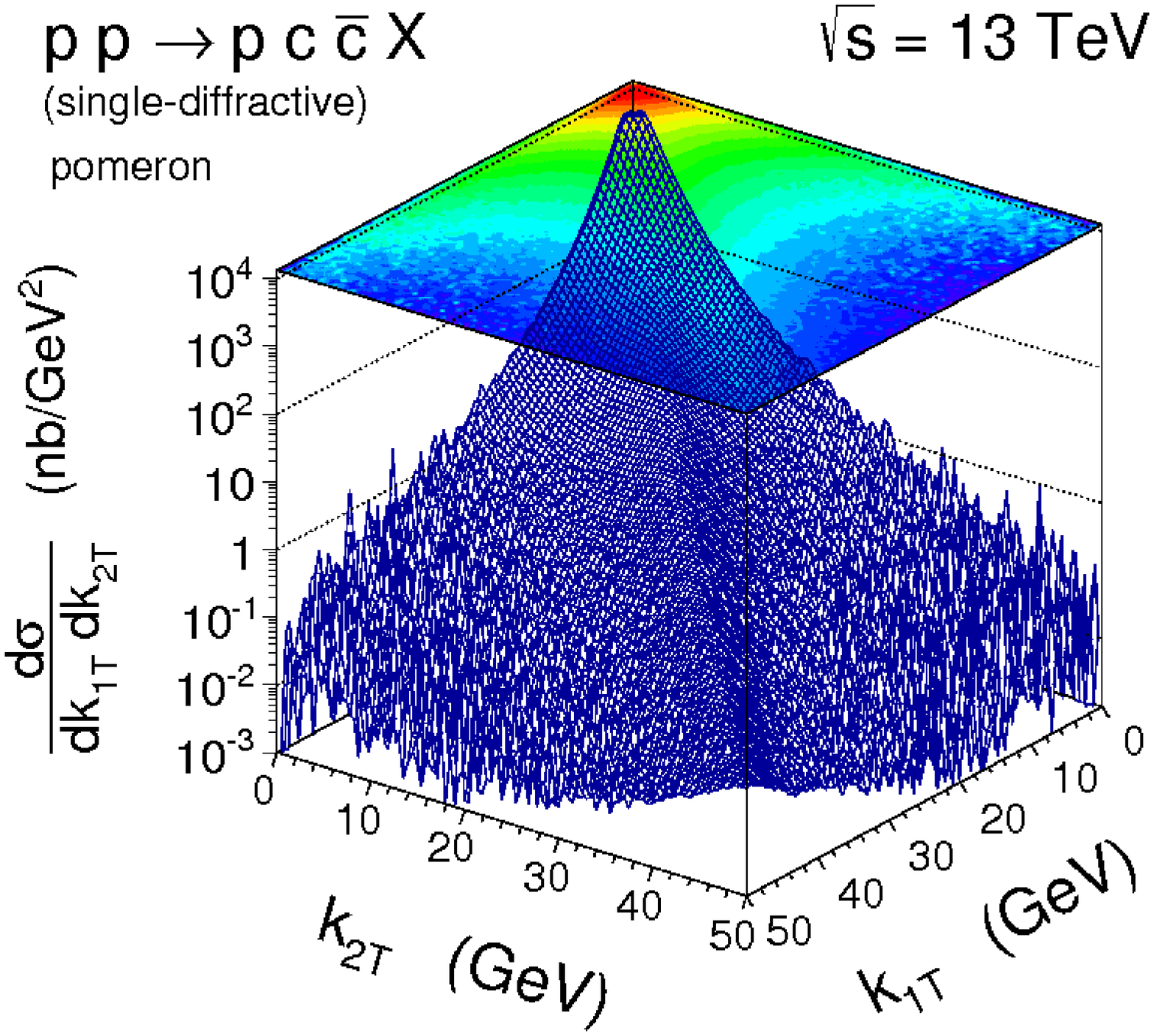}}
\end{minipage}
\hspace{0.5cm}
\begin{minipage}{0.4\textwidth}
 \centerline{\includegraphics[width=1.0\textwidth]{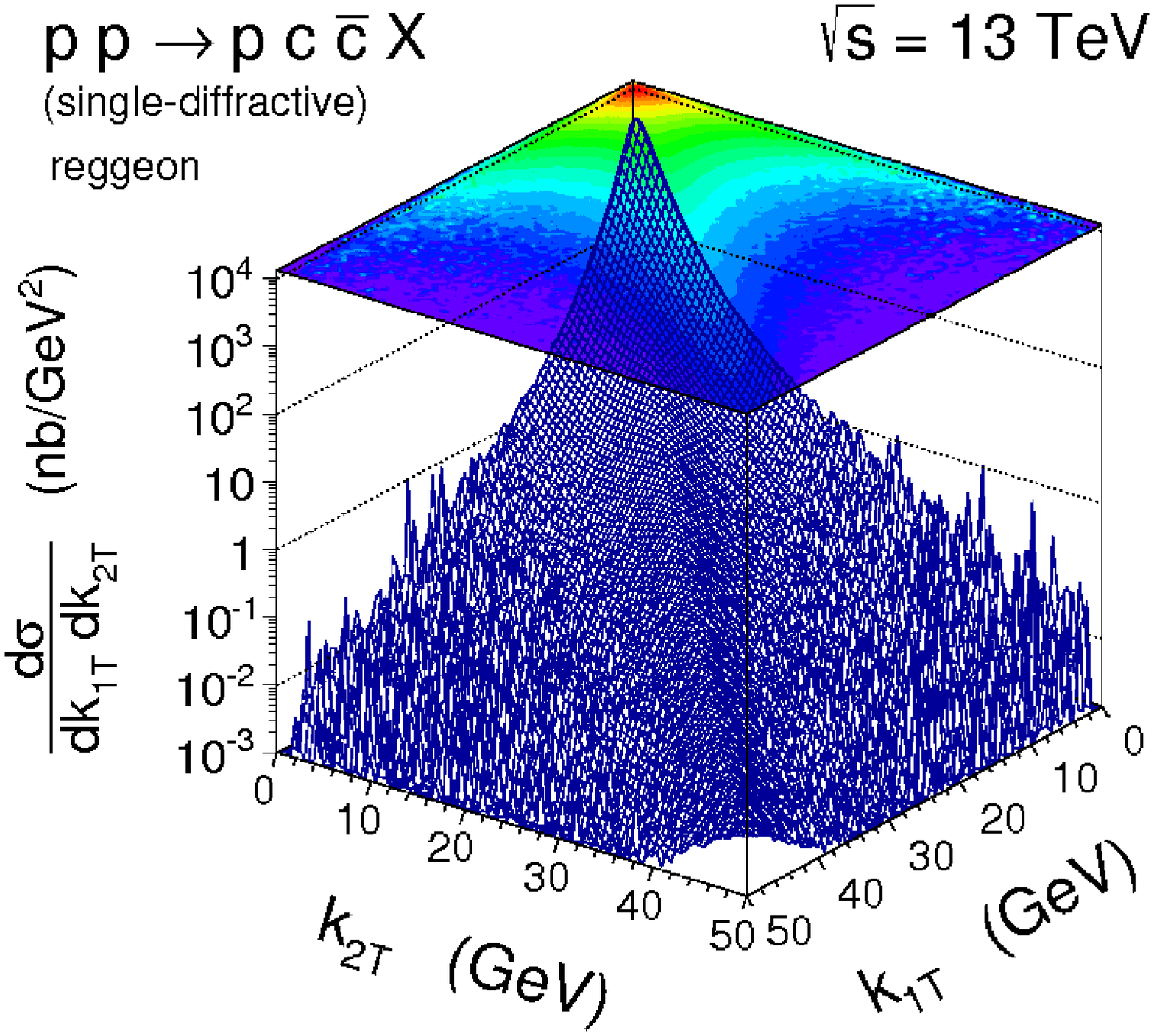}}
\end{minipage}
   \caption{
\small Double differential cross sections as a function of initial gluons transverse momenta $k_{1T}$ and $k_{2T}$ for single-diffractive production of charm at $\sqrt{s}=13$ TeV. The left and right panels correspond to the pomeron and reggeon exchange mechanisms, respectively. 
}
 \label{fig:q1tq2t_kT_PR}
\end{figure}
\begin{figure}[!htbp]
\begin{minipage}{0.4\textwidth}
 \centerline{\includegraphics[width=1.0\textwidth]{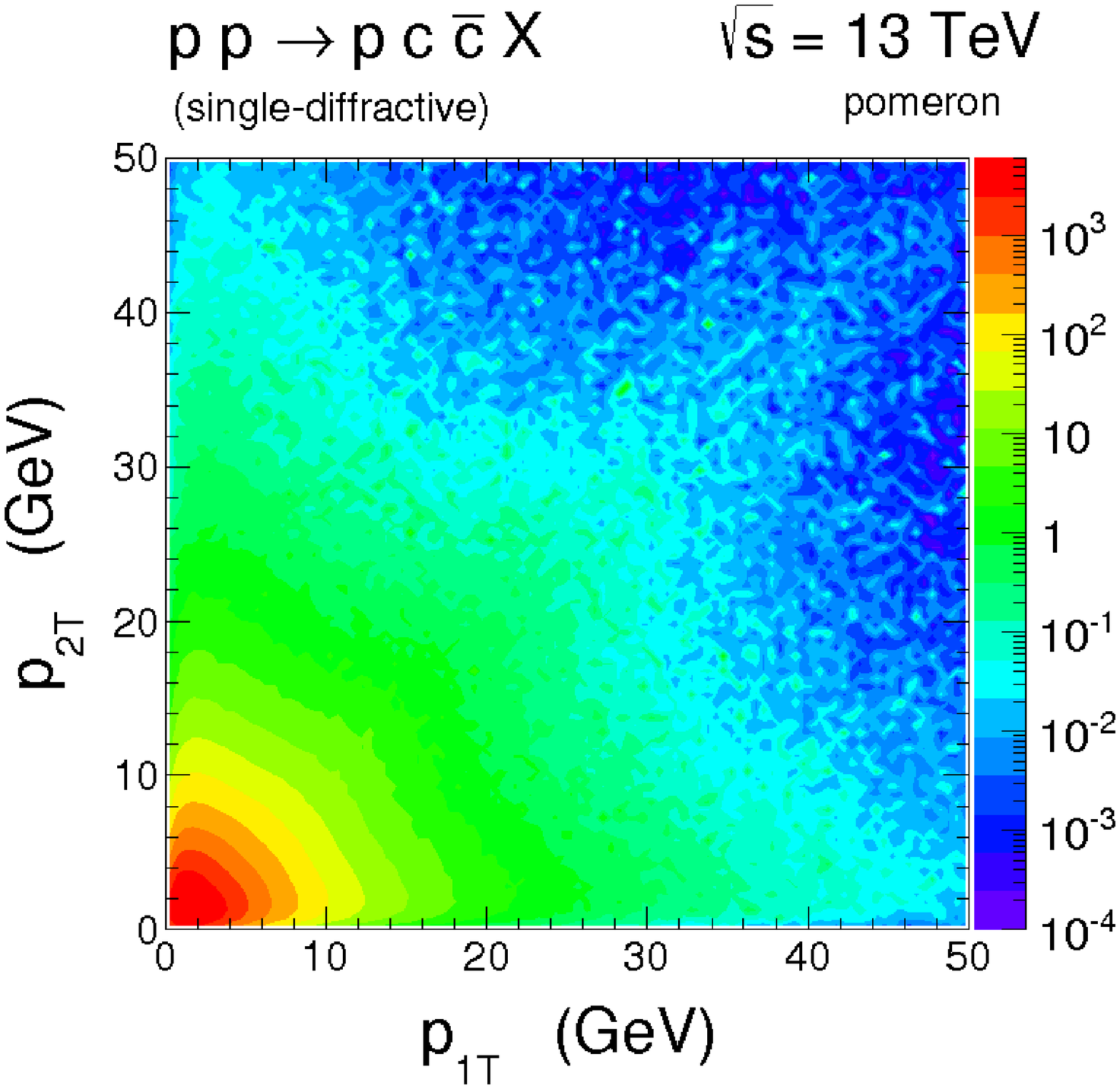}}
\end{minipage}
\hspace{0.5cm}
\begin{minipage}{0.4\textwidth}
 \centerline{\includegraphics[width=1.0\textwidth]{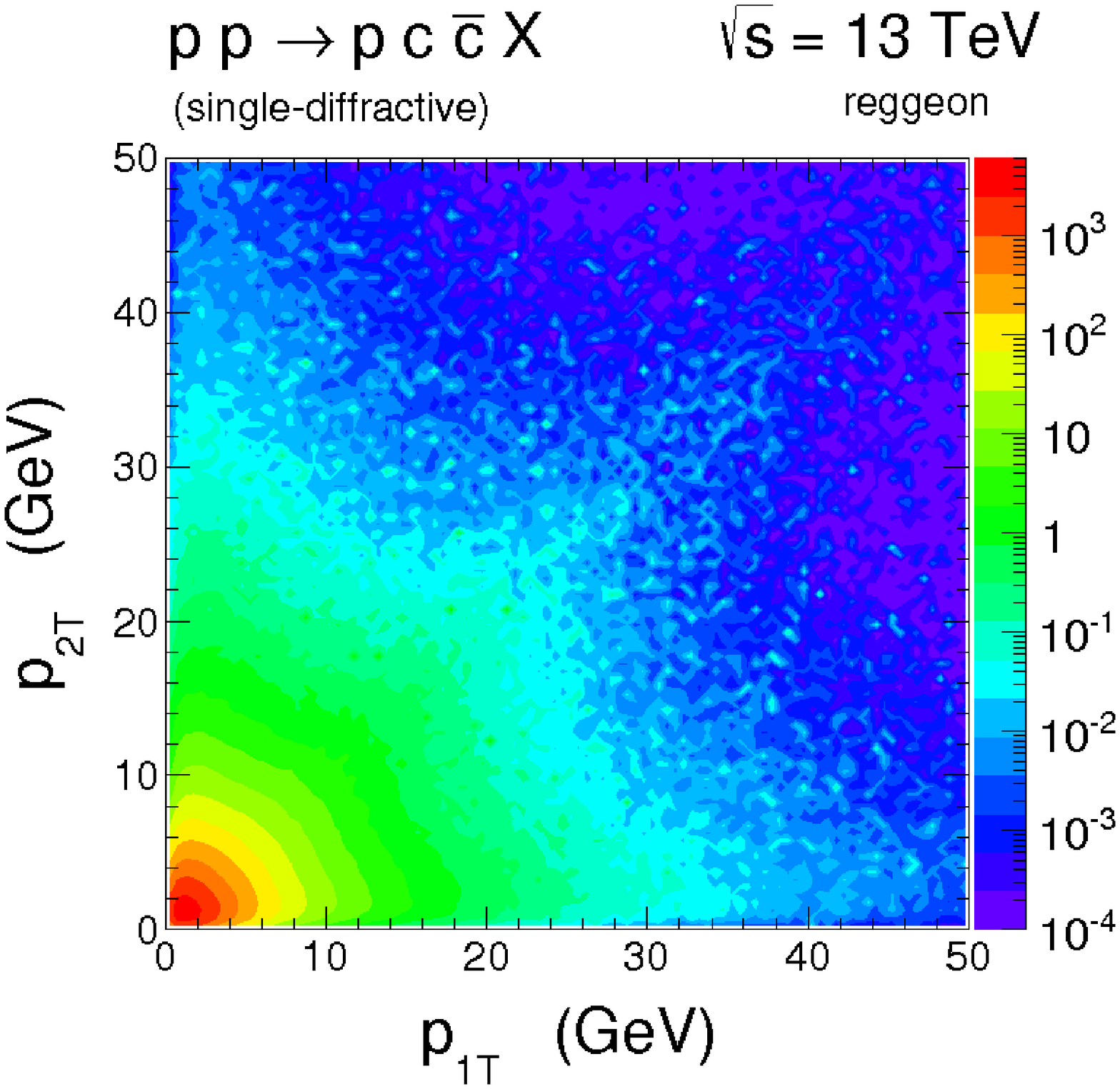}}
\end{minipage}
   \caption{
\small  Double differential cross sections as a function of transverse momenta of outgoing $c$ quark $p_{1T}$ and outgoing $\bar c$ antiquark $p_{2T}$ for single-diffractive production of charm at $\sqrt{s}=13$ TeV. The left and right panels correspond to the pomeron and reggeon exchange mechanisms, respectively. 
}
 \label{fig:p1tp2t_kT_PR}
\end{figure}

Figures~\ref{fig:q1tq2t_kT_PR} and \ref{fig:p1tp2t_kT_PR} show the
double differential cross sections as a functions of transverse momenta
of incoming gluons ($k_{1T}$ and $k_{2T}$) and transverse momenta of
outgoing $c$ and $\bar c$ quarks ($p_{1T}$ and $p_{2T}$), respectively. 
We observe quite large incident gluon transverse momenta. 
The major part of the cross section is concentrated in the region of
small $k_{t}$'s of both gluons but long tails are present. 
Transverse momenta of the outgoing particles are not balanced as they
were in the case of the LO collinear approximation.

\section{Conclusions}

Charm production is a good example where the higher-order effects are
very important. For the inclusive charm production we have shown that
these effects can be effectively included in the $k_t$-factorization approach \cite{Maciula:2013wg}. 
In our approach we decided to use the so-called KMR method to calculate
unintegrated diffractive gluon distribution (UDGD).
Having obtained the UDGD we have performed calculations of several
single-particle and correlation distributions. 
In general, the $k_t$-factorization approach leads to
larger cross sections. However, the $K$-factor is strongly dependent on phase space
point. Some correlation observables, like azimuthal angle correlation
between $c$ and $\bar c$, and $c \bar c$ pair transverse momentum
distributions were obtained in \cite{Luszczak:2016csq} for the first time.

\section*{Acknowledgments}
This work was partially supported by the Polish National Science Centre
grant DEC-2013/09/D/ST2/03724
as well as by the Centre for Innovation and Transfer of
Natural Sciences and Engineering Knowledge in Rzesz{\'o}w.



\begin{thebibliography}{100}
  
\bibitem{IS}
G.~Ingelman, P.E.~Schlein,
  Phys.\ Lett.\ B {\bf 152}, 256 (1985).
  
\bibitem{KMR2000}
  V.~A.~Khoze, A.~D.~Martin and M.~G.~Ryskin,
  Eur.\ Phys.\ J.\ C {\bf 18}, 167 (2000)
  [hep-ph/0007359].


\bibitem{Gotsman:2005rt}
  E.~Gotsman, E.~Levin, U.~Maor, E.~Naftali and A.~Prygarin,
  arXiv:0511060 [hep-ph].

\bibitem{Luszczak:2014cxa}
  M.~Luszczak, R.~Maciula and A.~Szczurek,
  Phys.\ Rev.\ D {\bf 91} (2015) no.5,  054024
  [arXiv:1412.3132 [hep-ph]].

\bibitem{Maciula:2013wg} 
  R.~Maciula and A.~Szczurek,
  Phys.\ Rev.\ D {\bf 87}, 094022 (2013)
  [arXiv:1301.3033 [hep-ph]].
  
\bibitem{Luszczak:2013cba}
  M.~Luszczak, W.~Schafer and A.~Szczurek,
  Phys.\ Lett.\ B {\bf 729} (2014) 15
  [arXiv:1305.4727 [hep-ph]].

\bibitem{Luszczak:2016csq}
  M.~Luszczak, R.~Maciula, A.~Szczurek and M.~Trzebinski,
  arXiv:1606.06528 [hep-ph].

\end{thebibliography}
\end{document}